\documentclass[aps,prb,twocolumn,nopacs,preprintnumbers]{revtex4}
	\usepackage{graphicx}
	\usepackage{bm}
	\usepackage{color}
	\usepackage{amsmath}
	\usepackage{epsfig}
	\usepackage{epstopdf}
	\usepackage{comment}

	\usepackage{mathrsfs}
	\usepackage{subfigure}

\begin{document}
	\title{A single nodal loop of accidental degeneracies in
	minimal symmetry:\\ triclinic CaAs$_3$}
	\author{Y. Quan}
	\affiliation{Department of Physics and Center for Advanced Quantum Studies, Beijing Normal University, Beijing 100875, China}
	\author{Z. P. Yin}
	\email[Corresponding author. Email address: ]{yinzhiping@bnu.edu.cn}
	\affiliation{Department of Physics and Center for Advanced Quantum Studies, Beijing Normal University, Beijing 100875, China}
	\author{W. E. Pickett}
		\email[Corresponding author. Email address: ]{pickett@physics.ucdavis.edu}
	\affiliation{Department of Physics,
	  University of California Davis, Davis CA 95616}
	\begin{abstract}
	The existence of closed loops of degeneracies in crystals has been intimately 
	connected to associated crystal symmetries, raising the question: what is the minimum 
	symmetry required for topological character, and can one find an example?  
	Triclinic CaAs$_3$, in space group $P{\bar 1}$ with only a 
	center of inversion, has been found to display, without need for tuning, a nodal 
	loop of accidental degeneracies with topological character, centered on one face 
	of the Brillouin zone that is otherwise fully gapped. The small loop 
	is very flat in energy, yet is cut four times by the Fermi energy, a condition that 
	results in an intricate repeated touching of inversion 
	related pairs of Fermi surfaces at Weyl points.  
	Spin-orbit coupling lifts the fourfold degeneracy along the loop, leaving
    trivial Kramers pairs.
	With its single nodal loop that emerges without protection from any point group
	symmetry, CaAs$_3$ represents the
	primal ``hydrogen atom'' of nodal loop systems.
	\end{abstract}
	\date{\today}
	\maketitle

	Nodal loop semimetals (NLSs) represent the most delicate type of topological phase
	in the sense that they arise from a closed loop of {\it accidental} degeneracies in
	the Brillouin zone. In some ways they complement the topological character of
	Weyl semimetals\cite{wan2011} by displaying surface Fermi arcs or Fermi lines,
	or both. Several structural classes of NLSs have been identified, always
	associated with specific space group symmetries that enable, or in common 
	parlance protect, the necessary degeneracies. On the other hand, the early
	theoretical work\cite{Herring1937,HerringThesis} presumed only the minimum 
	symmetry necessary to allow a
	nodal loop: time reversal symmetry and a center of inversion. This limiting
	case of ``minimal symmetry'' has prompted us to look for an example and
	understand its behavior.

	When the little group at wavevector $\vec k$ contains only the identity, the Hamiltonian 
	$H(\vec k)$ has matrix elements between states with neighboring eigenvalues and anticrossings occur as some parameter of H is varied. von Neumann and Wigner first investigated the conditions under which degeneracies nevertheless occur, so-called accidental degeneracies,\cite{Neumann1929} where matrix elements vanish for no physical reason. Herring generalized their arguments to accidental degeneracies in three dimensional (3D) crystals.\cite{Herring1937, HerringThesis} with some extension by Blount.\cite{ Blount1962} Herring pointed out, for example, that a mirror plane provides a natural platform for a ring of degeneracies. If a band with even mirror symmetry is higher in energy than a band of odd symmetry at $\vec k_1$ but lower at $\vec k_2$ (both on the mirror plane), then due to the continuity of eigenvalues and differing symmetry, on any path connecting them there must be a point of degeneracy. The locus of such degeneracies maps out either a loop encircling one of the points, or an extended line from zone to zone separating the two points (which, considering periodicity, also becomes a closed loop topologically). 

	The topologically singular nature of such nodal loops was established by Berry.\cite{Berry1985} Allen demonstrated\cite{Allen2007} how, with minimal symmetry available, these loops of degeneracies are destroyed by spin-orbit coupling (SOC). Special symmetries can
	enable nodal loops in the presence of SOC, for example a screw axis in the example of Fang
	{\it et al.}\cite{Fang2015}   Burkov {\it et al.} made the modern rediscovery of nodal loops
	and illustrated the type of Weyl-point connected electron and hole Fermi surface that should be expected when the band energies around the loop cross the Fermi energy.\cite{Burkov2011} 
	Such nodal loops should be common, and indeed have been found even in high symmetry elemental metals.\cite{hirayama2016} Nodal loop semimetals based on crystal symmetries, especially mirror symmetries, have appeared in several models\cite{Burkov2011,Phillips2014,Kim2015,Fang2015,Heikkila2015,Mullen2015} and crystal structures.\cite{Pardo2010,Fang2012,Yu2015, Huang2015, Xu2015, Lv2015, Shekhar2015,  Weng2015, Ahn2015, Sun2015, Yang2015, Neupane2016, Huang2016, Zhao2015} 

	Before the discussion of Burkov {\it et al.}\cite{Burkov2011}, a nodal 
	loop of a pair of coinciding Fermi rings -- a nodal ring coinciding with the 
	Fermi energy $E_F$ -- had been discovered in calculations of a ferromagnetic compensated semimetal 
	SrVO$_3$ quantum confined within insulating SrTiO$_3$.\cite{Pardo2010} Mirror symmetry 
	was a central feature in providing compensation and the degenerate nodal loop
	coinciding with $E_F$. 
	What is unlikely but not statistically improbable is: (1) having the nodal loop cut by 
	$E_F$  while (2) the remainder of the Brillouin zone is gapped. Such loops will 
	have real impact, and possible applications, when they are the sole bands around  
	$E_F$, because they generate topological character with corresponding 
	boundary Fermi arcs or points at zero energy.

	\begin{figure}[!ht]
	 \includegraphics[width=0.8\columnwidth]{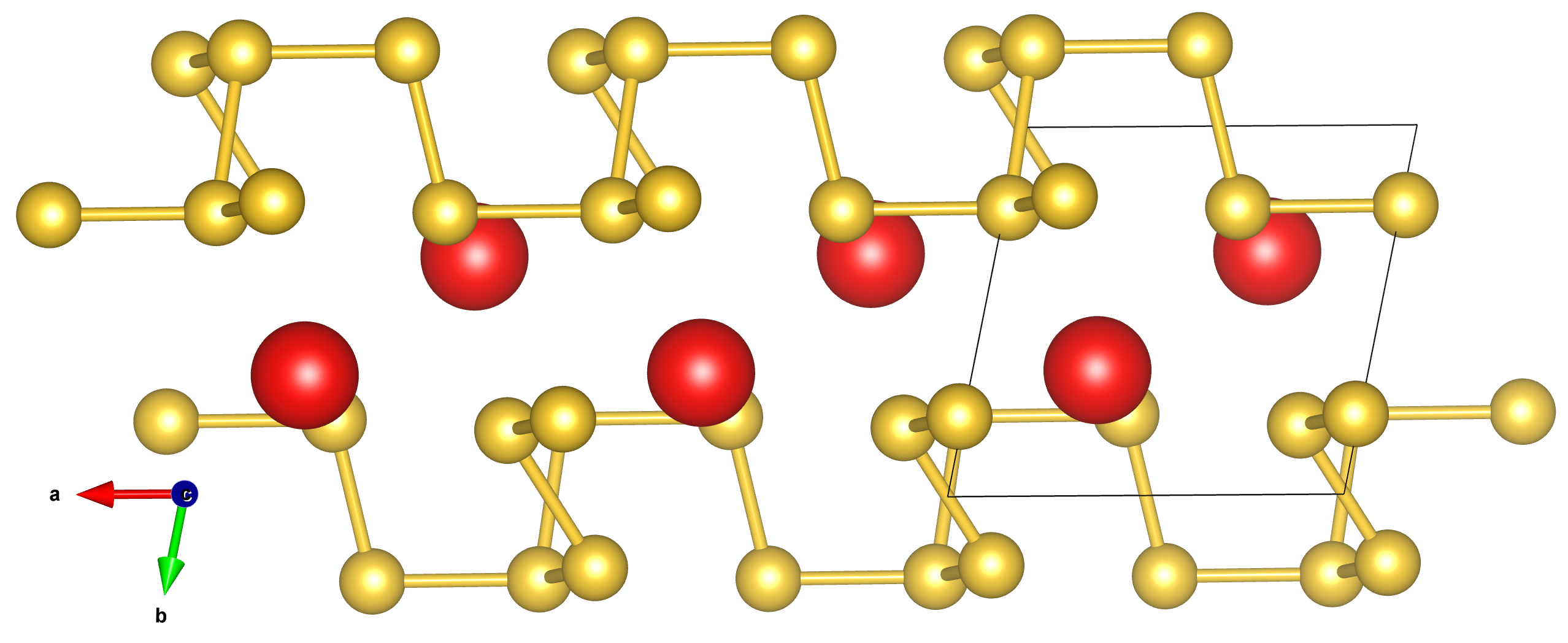}
	  \caption{Crystal structure of CaAs$_3$, viewed in the $b$-$c$ plane. Arsenic atoms (yellow) form two-dimensional chains similar to black phosphorus. The center of inversion lies midway between neighboring Ca ions (shown in red).}
	  \label{struct}
	\end{figure}

	Among his several results relating crystal symmetries to accidental degeneracies 
	without consideration of SOC,
    Herring\cite{Herring1937, HerringThesis} 
	found that inversion symmetry 
	${\cal P}$ alone is sufficient to allow nodal loops of degeneracies (fourfold: two
	orbitals times two spins), a result extended 
	recently.\cite{Burkov2011,Fang2015,Kim2015} Simply stated, ${\cal P}$ symmetry
	leads to a real Bloch Hamiltonian $H(\vec k)$ if the center of inversion is taken as the origin. The minimal (for each spin)
	2$\times$2 Hamiltonian then has the form $H(\vec k)=f_k\tau_x + g_k\tau_z$ (neglecting 
	spin degeneracy for the moment) with real functions $f_k, g_k$; $\vec \tau$ represents 
  the Pauli matrices in orbital space. Degeneracy of the eigenvalues 
  $\varepsilon_k = \pm(f_k^2 + g_k^2)^{1/2}$ requires $f_k=0=g_k$, two conditions 
  on the 3D vector $\vec k$, giving implicitly (say) $k_y={\cal K}(k_x,k_z)$ for 
  some function ${\cal K}$. This condition either has no solution, or else corresponds 
  to a loop ${\cal L}$ of degeneracies. Allen has given a constructive 
  prescription\cite{Allen2007} for following the nodal loop once a degeneracy is located. 

	Any such loop will not lie at a single energy,\cite{Herring1937, Allen2007, Burkov2011} 
	and as mentioned only acquires impact when the dispersion around the loop crosses $E_F$,
	with a gap elsewhere. 
	This intersection results in a pair (or an even number) of points where, in the absence 
	of spin-orbit coupling, the valence and conduction band Fermi surfaces touch. 
	The dispersion at the Fermi contact points will, barring accidents of zero probability, 
	be massless in all three directions\cite{HerringThesis} -- Weyl points. Thus at this level 
	(before SOC) the nodal loop semimetal is a special subclass of 3D Weyl semimetal. 

	Topics that have not been addressed are: how little symmetry is necessary for topological character to be retained, what are the consequences, and can an example with minimum symmetry be found? 
	The line of reasoning above applied to the case of no inversion center ({\it i.e.} no crystal symmetry at all) dictates that all of the coefficients of $\sigma_x, \sigma_y, \sigma_z$ in $H(\vec k)$ vanish.
	Accidental point degeneracies are thus possible by tuning, while a line of degeneracies occurs with zero probability. 

	Discovery and study of topological nodal line semimetals protected by crystal symmetry is developing rapidly.\cite{hirayama2016,Yu2015,Weng2015,Ahn2015,Yang2015}  The class $TPn$ ($T$=Nb, Ta; $Pn$=P, As) lacks an inversion center but contains several crystalline symmetries facilitating nodal loops.\cite{Huang2015, Xu2015, Lv2015, Shekhar2015, Weng2015, Ahn2015, Sun2015, Yang2015} The cubic antiperovskite Cu$_3$PdN contains nodal loops in a background of metallic bands,\cite{Kim2015, Yu2015} the BaTaSe$_4$ family has nodal loops in its band structure enabled by symmetry, and as mentioned cubic elemental metals contain loops within their metallic bands.\cite{hirayama2016} Here we show that triclinic CaAs$_3$ is an example of a minimal symmetry nodal loop semimetal with a single loop of degeneracies, providing the ``hydrogen atom'' of the class of nodal semimetals..

	\begin{figure}[!ht]
	  \includegraphics[width=\columnwidth]{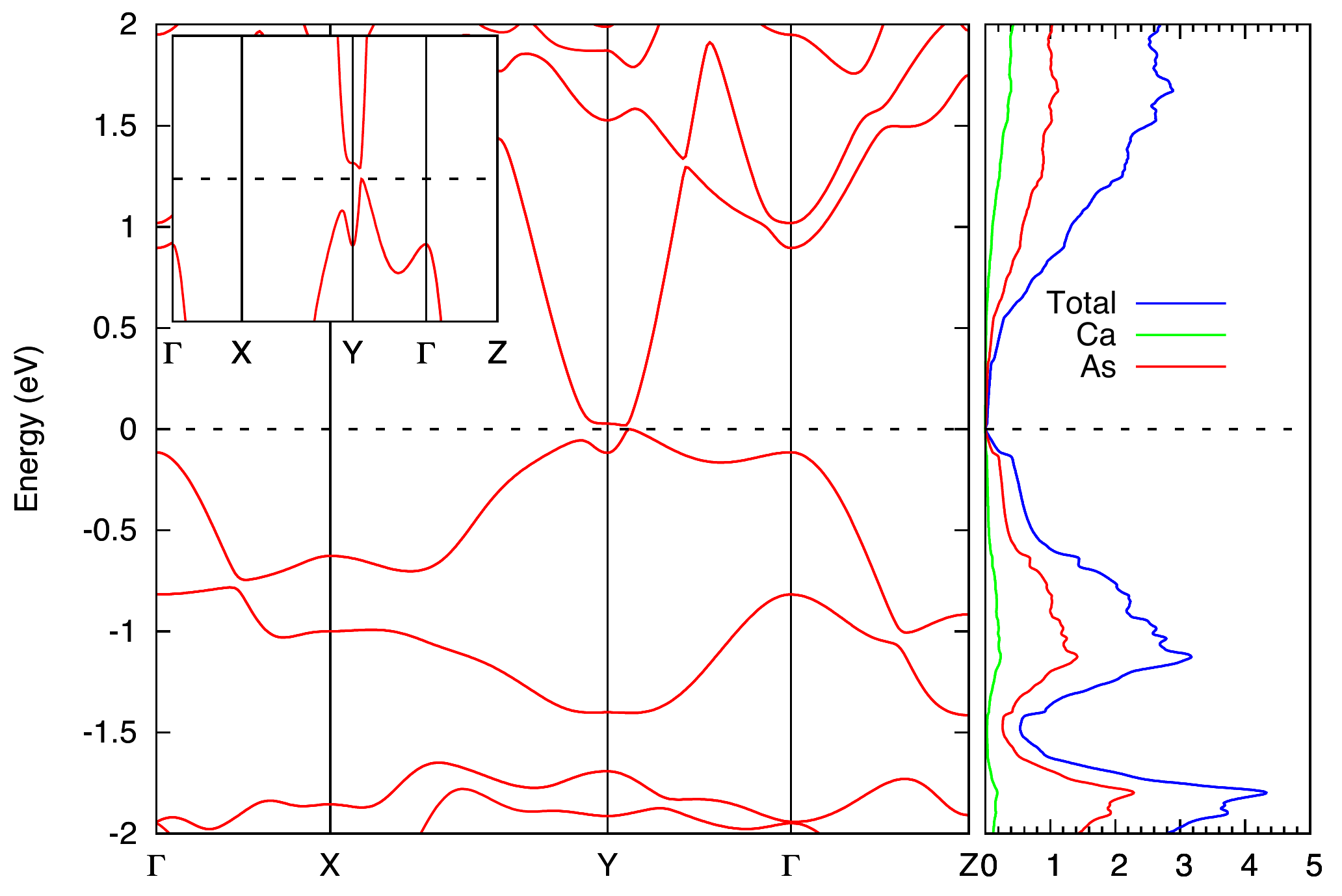}
	  \caption{Band structure of CaAs$_3$ along a few special direction, from a GGA+mBJ+SOC calculation, and (right panel) the density of states.The region of interest lies near the Y=$\vec b^*/2$ zone boundary (ISIM) point. Band inversion at $Y$ can be easily imagined by ignoring the mixing that causes anticrossing along the $X-Y$ direction. Even without SOC, a gap of $\sim10$ meV separates occupied and unoccupied states along the $Y$-$\Gamma$ direction (see inset).}
	\label{bands}
	\end{figure}

	CaAs$_3$ and three isovalent tri-arsenides (Ca$\rightarrow$Sr, Ba, Eu) were synthesized more than thirty years ago, with their structure, transport, and optical properties studied by Bauhofer and collaborators.\cite{bauhofer1981,Oles1981} CaAs$_3$ is the sole triclinic member of this family, with space group $P\bar{1}$ (\#2) containing only an inversion center, lying midway between Ca sites.\cite{bauhofer1981} Heavily twinned samples of CaAs$_3$ has been reported as insulating in transport measurements\cite{bauhofer1981} but curiously display\cite{Oles1981} in far infrared reflectivity a Drude weight corresponding to 10$^{17}$-10$^{18}$ carriers per cm$^3$. 

	The sole symmetry condition in $P\bar{1}$ symmetry on the energy bands is $\varepsilon_{-k}=\varepsilon_k$. This simplicity indicates that ``symmetry lines'' are simply convenient lines with a trivial little group. $P{\bar 1}$ symmetry does however provide eight inversion symmetry invariant momenta (ISIM) $m\frac{a^*}{2}+n\frac{b^*}{2}+p\frac{c^*}{2}, m,n,p=0,1,$ in terms of the primitive reciprocal lattice vectors $a^*, b^*, c^*$. At these ISIMs, which are the analog of (and equivalent to) the time reversal invariant momenta (TRIMs) important in topological insulator theory,\cite{Teo2008} eigenstates have even or odd parity. Isolated nodal loops either (a) must be centered at an ISIM, or (b) they occur in inversion related pairs. Due to the low symmetry, finding unusual characteristics (viz. the occurrence of and center of a nodal loop)  necessitates meticulously searching in band inversion regions. 

	The linearized augmented plane wave method as implemented in WIEN2k\cite{wien2k} was applied with the generalized gradient approximation (GGA) exchange-correlation potential.\cite{GGA}  $R_{m}K_{max}$=7 is a sufficient cutoff for the basis function expansion in this $sp$ electron material. Studies have shown that GGA may underestimate relative positioning of valence and conduction bands in semiconductors and semimetals, and that the modified Becke-Johnson (mBJ) potential provides a reasonably accurate correction.\cite{zhang} Thus we rely on the GGA+mBJ combination throughout.
	The impact of the As SOC is assessed.

	The CaAs$_3$ band structure and density of states (DOS) in directions along reciprocal
	lattice vectors and in the energy range from -2 eV to 2 eV, shown in Fig.~\ref{bands}, suggests small-gap insulating character. 
	Valence and conduction bands are separated in energy except for an evident band inversion 
	at the $Y\equiv \vec b^* /2$ zone boundary ISIM point. 
	Note that with non-ISIM points having a trivial little group, bands do not cross except 
	at accidental degeneracies, and these will coincide with any given line with zero 
	probability.\cite{Allen2007} 
	The combination of ${\cal P}$ symmetry and periodicity is enough to ensure that band 
	energies at $\vec b^*/2 \pm (0,\delta k_y,0)$ are equal, thus (relative) band extrema 
	occur at the ISIMs, and can be observed at $X, Y,$ $Z$, and $\Gamma$ in Fig.~\ref{bands}. 

	\begin{figure}[!ht]
	  \includegraphics[width=\columnwidth]{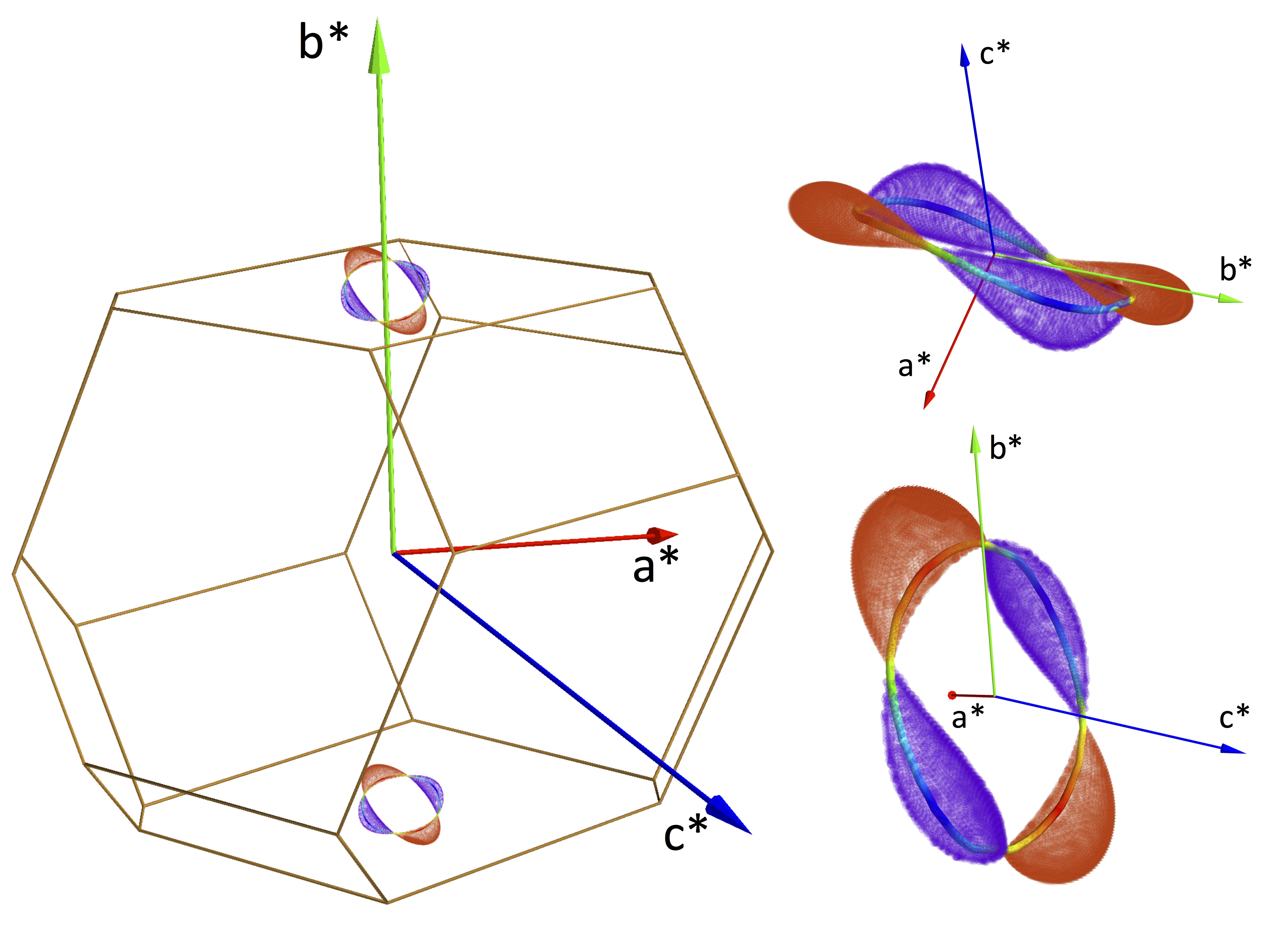}
	  \caption{Left panel: Brillouin zone of CaAs$_3$, showing 
	the nodal loop centered at $Y$ on the top (and bottom) face 
	of this view of the zone. 
	  Right panels: two perspective views of the nodal line 
	  enclosed within the Fermi surfaces, with electron and hole
	  surfaces denoted by different colors. The notations  $a^*, b^*, c^*$
	  of the reciprocal lattice vectors denote direction only.}
	  \label{NodalLoop-FSs}
	\end{figure}

	Searching the band inversion region, a loop ${\cal L}$ of accidental degeneracies 
	centered at $Y$ was mapped out, {\it i.e.} there is no gap. Its position in the BZ is shown in Fig.~\ref{NodalLoop-FSs} together with two perspective views of the Fermi surfaces (FSs). The loop, resembling a nearly planar lariat, is cut by $E_F$ at not two but {\it four} points, each point being a touching point for a hole and electron FS (guaranteed by the nodal degeneracies).  At this level (no SOC) the spectrum is that of a semimetal with FSs touching at four Weyl points. 
	The loop energy lies in the -20 meV to +20 meV range, making it a very flat 
	nodal loop in the energy domain as well as in momentum space. 

	 The surface Fermi arcs of a few 3D Weyl semimetals are now well studied.\cite{wan2011} 
	The analogous states in NLSs were discussed originally by 
	Burkov {\it et al.}\cite{Burkov2011} Projected onto a surface, ${\cal L}$
	will enclose an area (which we call a ``patch'') 
	within which topologically-required surface states (``drumhead states'') reside. 
    For CaAs$_3$, ${\cal L}$ projected along $a^*$ leaves a roughly circular
    patch, and along $c^*$ 
    roughly elliptical, consistent with what can be surmised from 
    Fig.~\ref{NodalLoop-FSs}.  
    The $b^*$ axis however lies nearly within the plane of the loop, projecting
    to a very slender patch.
	A plot along a $\vec k$-line crossing the patch will reveal a 
	surface band starting at 
	the edge of this patch and ending when the k-line leaves the patch. Considering the constant energy contours (potential Fermi lines) in the patch, they may be closed lines or isolated 
	arcs that terminate at the boundary of the patch. 

\begin{figure}[!ht]
	\includegraphics[width=\columnwidth]{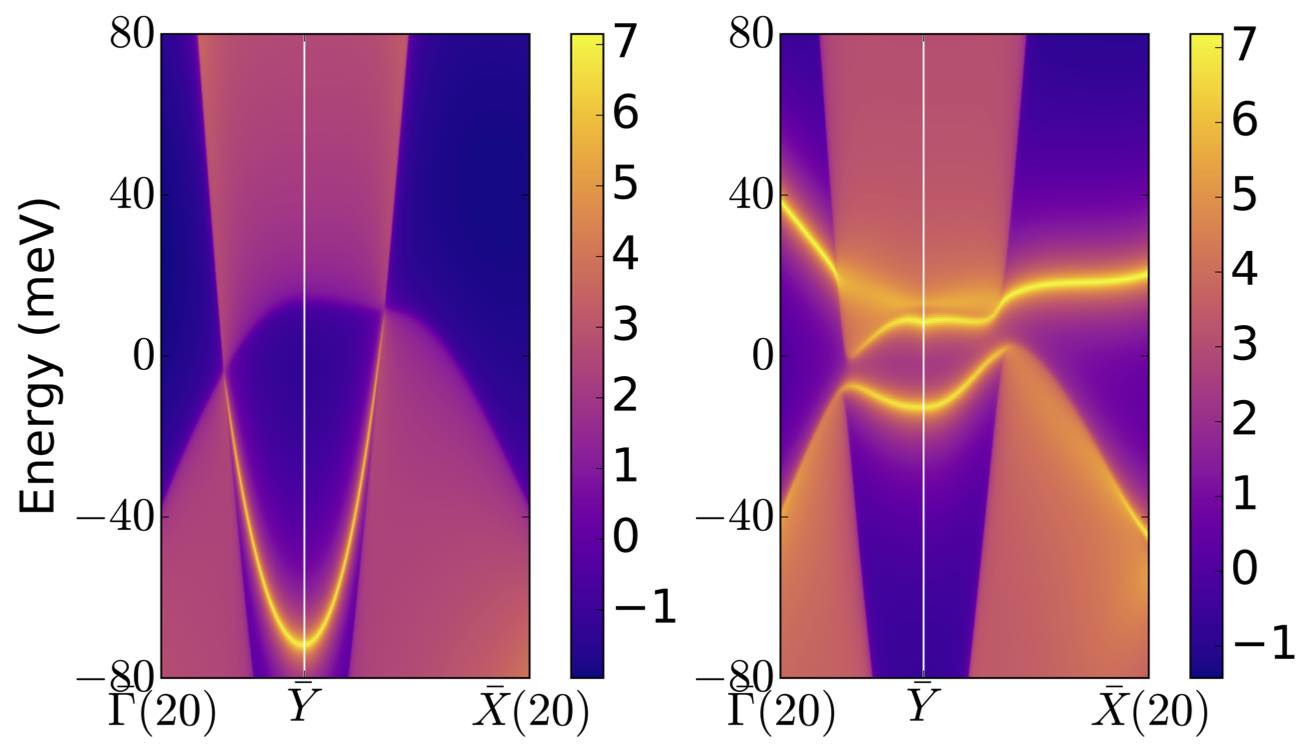}
	\caption{Edge states (peaks in the spectral density) of CaAs$_3$ calculated 
	using the MLWF tight binding representation truncated at the (001) surface. 
	The panels compare spectra before (left) and after (right) inclusion of SOC. 
	SOC hardly affects the projected bulk  bands while altering the surface
	(bright color) bands strongly.
	The notation ``X(20)'' for example, indicates the end point is 20\% of the 
	distance toward X.}
	\label{3SurfStates}
	\end{figure}

	Surface band plots along special directions
	$\bar{Y}-\bar{\Gamma}-\bar{X}$ are shown for 
	the (001) surface in Fig.~\ref{3SurfStates}. 
	As mentioned, the Fermi energy cuts the nodal loop, hence it intersects 
	the surface patch band resulting in one or more Fermi lines on each surface. 
    The surface band disperses along these lines shown by 70 meV.
	We have confirmed other studies\cite{Zhao2015,Pi2016} that indicate that surface
	bands obtained from Wannierization followed by truncation to obtain a
	surface can be sensitive to numerical procedures and the chosen surface
	termination, so these bands are not a definitive prediction of the physical
	surface states. Moreover, non-topological surface bands such as from 
	dangling bonds may appear as well.

	{\it Effect of spin-orbit coupling.} 
Allen demonstrated in generality the effect of SOC on the 
nodal loop, using a two band
model in the low energy regime.\cite{Allen2007}
Without any symmetry operation to cause
the SOC matrix element to vanish, which is the case in CaAs$_3$, the
degeneracy is opened to a $k$-dependent gap $\xi_k$ along the entire loop,
which retains an inactive Kramers degeneracy.
For integration around a circuit surrounding the loop ${\cal L}$, the
topological phase $\pm\pi$ is replaced by a non-topological Berry phase
that is dependent on the radius of the circuit.
A magnetic field coupled to spins splits the Kramers degeneracies everywhere,
giving four distinct bands near ${\cal L}$.
Allen's paper should be consulted for specific dependencies on the
materials parameters.

	The SOC splitting of the atomic As $4p$ level is 270 meV. Since the 
	bands that are inverted at $Y$ are primarily As $4p$ character, the 
	SOC-driven band shifts will be some appreciable fraction of this value. 
	Given the 40 meV span in energy of the nodal loop, large enough SOC can  
	open a gap. 
	The bulk band projection, visible in Fig.~\ref{3SurfStates},  is 
	altered little by SOC.
	Within the accuracy of the Wannier interpolation and surface projection, the
	result is characteristic of separated valence and conduction bands that however
	leave little or no gap.

	Fig. ~\ref{3SurfStates} reveals that the surface spectrum evolves considerably 
	under SOC. Most evidently, the dispersion of the valence (occupied) surface
	band has decreased from 70 meV to only 10
	meV.  
	If SOC coupling is large enough compared to the dispersion around ${\cal L}$, 
	the system will be gapped by SOC, and CaAs$_3$ seems on the borderline of
	this situation. If a gap opens, it may provide  a distinct topological
     character, signaled by
     the usual $\nu_0(\nu_1 \nu_2 \nu_3)$ indices.
    We calculate that CaAs$_3$, with SOC
    taken into account, is a topological phase with indices 
	$\nu_0(\nu_1\nu_2\nu_3)$=1(010) using the criteria of Fu and Kane. 

         The spectrum in the right hand panel of Fig.~4 indicates
    the surface bands that will be topological insulator boundary states
    if SOC is large enough to give a gap. Otherwise they are topological
    surface states of a semimetal arising from indirect overlap, that is,
    a topological semimetal neither Weyl nor nodal loop. Our results in
    Fig.~4 indicate that CaAs$_3$ is extremely close to the topological
    semimetal - topological insulator transition. 

\begin{figure}[!ht]
	\includegraphics[width=0.48\columnwidth]{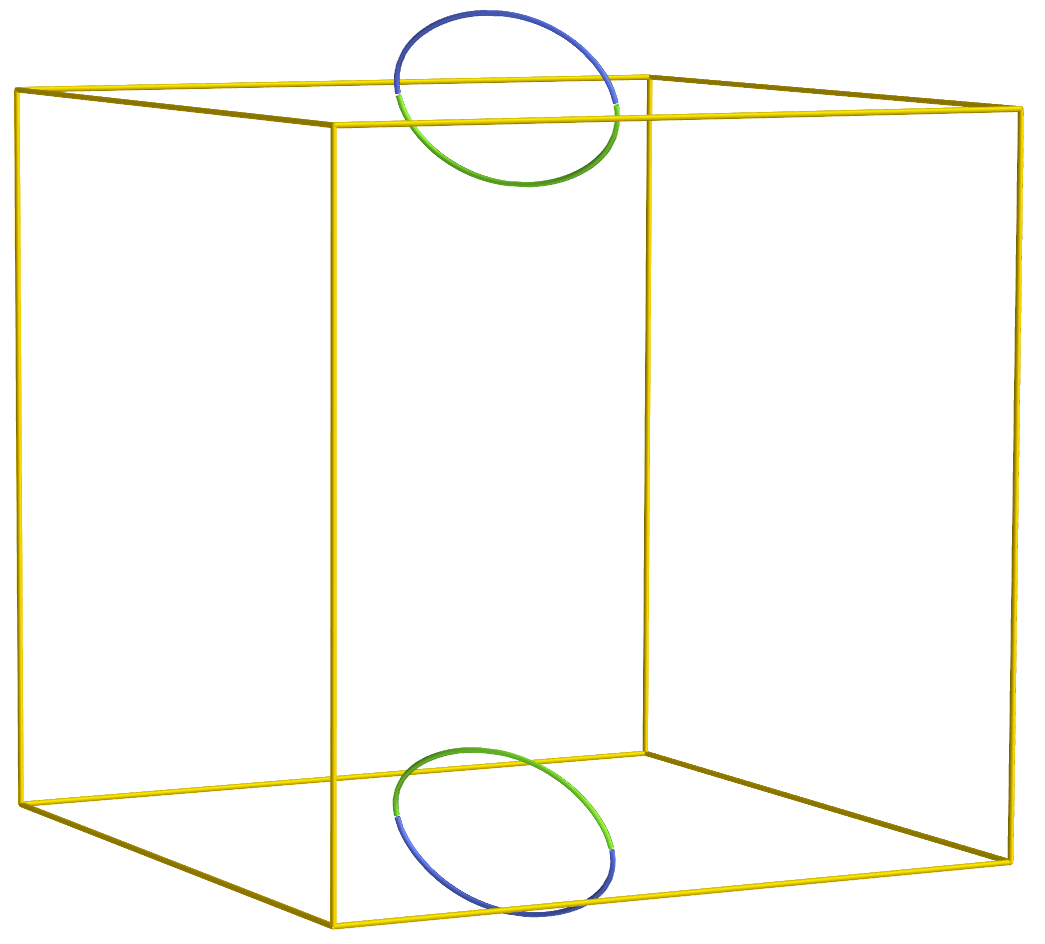}
	\includegraphics[width=0.48\columnwidth]{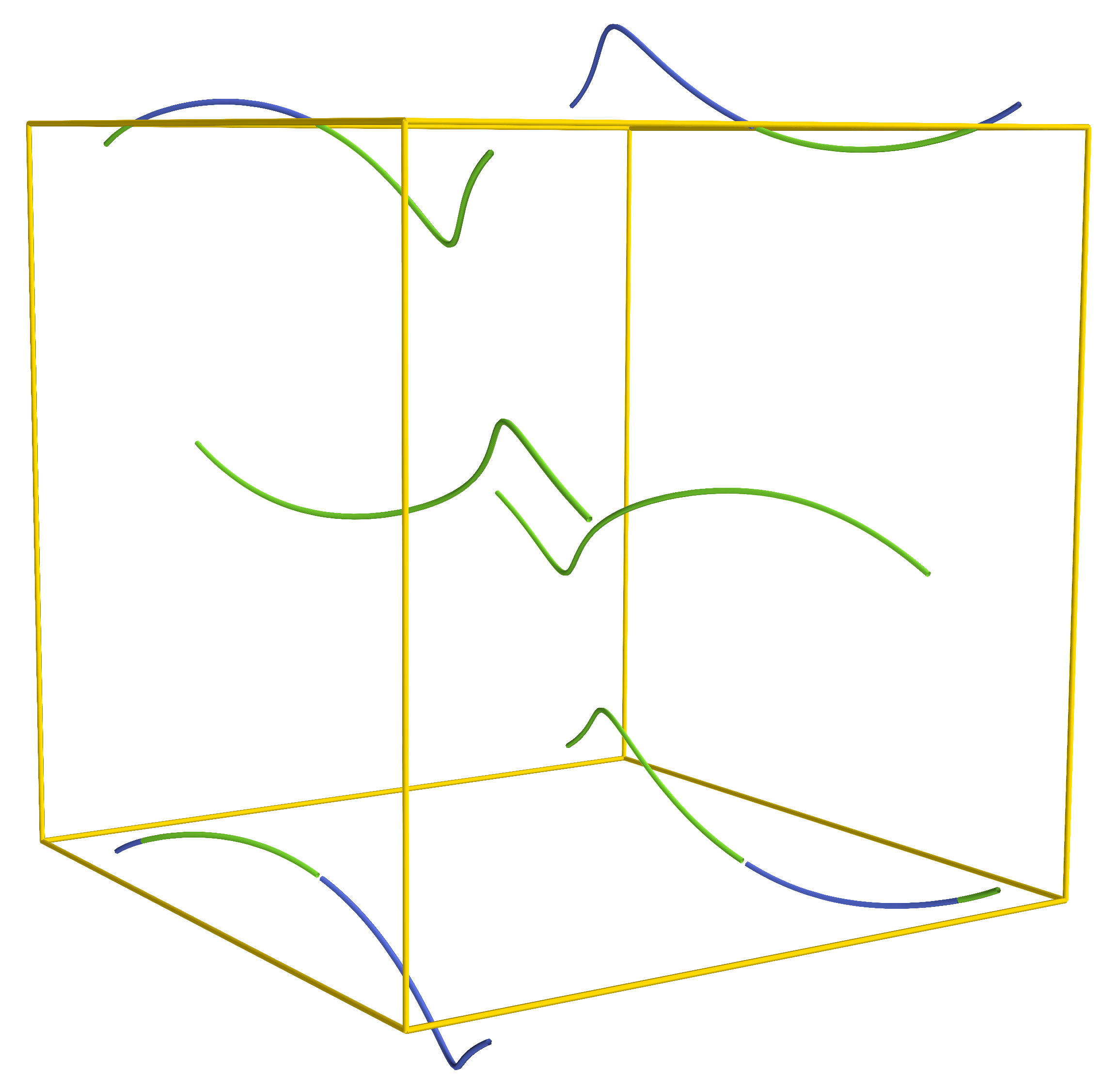}
	\caption{Nodal lines of accidental degeneracies for the model Hamiltonian' the dots indicate
	where a loop passes into a neighboring Brillouin zone. 
	For $m$=1.44 on the left, a single loop is centered on the ISIM point $\frac{\vec b^*}{2}$. 
	The $m$=0 case is shown on the right, with two pairs of inversion symmetry related lines threading from zone to zone. 
	}
	\label{loop}
	\end{figure}

	{\it Topological nodal loop from an effective Hamiltonian.}
	The band structure near E$_F$ of CaAs$_3$, with the highest 
	valence band inverted across the lowest conduction band at $Y,$ 
	was fit to a tight-binding
	model. Away from $Y$ CaAs$_3$ is gapped, making this compound ideal for 
	observing a topological nodal line. For simplicity one can imagine 
	the crystal deformed by an affine transformation to have orthogonal 
	axes with $a$=$b$=$c$=$1$. 
    We consider the following two orbital, 
    non-inversion symmetric Hamiltonian which reproduces 
	the essential features of the inverted band region of CaAs$_3$. 
	It includes nearest neighbor hopping between like orbitals 
	$\{t_{\alpha},\alpha=1-3\}$, and between unlike orbitals 
	$\{t_{\alpha},\alpha=4-6\}$ having differing parity:
	\begin{eqnarray*}
\tilde{H}(\vec{k})& = &g_{\vec k}\tau_z+f_{\vec k}\tau_x +\xi\sigma_z \tau_y\\ \nonumber
    f(k_a,k_b,k_c)&=&t_4 \sin k_a + t_5 \sin k_b + t_6 \sin k_c  \\ \nonumber
    g(k_a,k_b,k_c)&=&m-t_1\cos k_a - t_2\cos k_b-t_3\cos k_c,
	\end{eqnarray*}
 where \{$\tau_j$\},\{$\sigma_j$\} are the 2$\times 2$ matrices
   in orbital and spin space respectively, and $\xi$ is the SOC parameter. 
	This Hamiltonian describes two particle-hole symmetric bands 
 $\pm |g_k|$ with centers separated by $2|m|$, coupled by $f_k$, and including
 intra-orbital SOC, 
 with eigenenergies $\varepsilon_{k,\pm}=
  \pm \sqrt{g^2_{\vec k}+f^2_{\vec k} +\xi^2}$. 
 Evidently SOC ($\xi \neq 0$) 
 splits the degeneracy everywhere.\cite{Allen2007,Fang2015} 
 To mimic CaAs$_3$ we consider the site energy $m$ and hopping parameters 
 (in eV) $m=1.64$, $t_1=0.37$, $t_2=-0.95$, $t_3=0.37$, $t_4=-0.18$, $t_5=0.12$, $t_6=0.38$. 
	Without SOC ($\xi=0$), degeneracy $f_k = 0 = g_k$ is realized around a nodal loop 
    flat in energy (at zero energy). The loop, centered 
	at $Y$ but otherwise depending on parameters, and shown in the left panel of 
    Fig.~\ref{loop}, resembles the nodal 
	loop of CaAs$_3$ pictured in Fig.~\ref{NodalLoop-FSs}. 

	The evolution of the loop topology can be followed by varying the 
	band separation $2m$. Two types of lines of accidental degeneracies 
	may emerge from the Hamiltonian: a closed nodal loop as in CaAs$_3$, 
	or a line extending from zone to zone, which by zone periodicity 
	become closed lines on the 3D-torus, the difference from the former
	being that they must occur in pairs. In Fig. ~\ref{loop}, the two 
	types of loops are plotted in the first Brillouin zone. On the left, 
	where $m$=1.44, a single loop is centered at $Y$. 

    Varying $m$ tunes 
	the system through an evolution from an odd number (one) to an
	even number (four) of nodal loops. 
	The right panel in Fig.~\ref{loop} ($m$=0)
	has two pairs of inversion symmetric nodal loops threading 
	through extended Brillouin zones. 
 Because the loop is flat in energy (at zero), adding
SOC immediately opens a global gap  of $2\xi$. 
We find the resulting state to
have indices 0(000), {\it i.e.} 
a trivial insulator,  A different, lower symmetry model would be
necessary to produce a topological insulating state such as occurs
in 1(010) CaAs$_3$.

	In this work we have studied the electronic and topological properties of 
	triclinic CaAs$_3$, which is distinguished by possessing the lowest possible 
	symmetry for a nodal loop semimetal.  In the absence of spin-orbit coupling, 
	CaAs$_3$ has a single nodal loop (others have loops occurring in pairs)
	that is cut by the Fermi level four times. Spin-orbit coupling leads not 
	only to lifting of the nodal loop degeneracies and separation of valence
	and conduction bands complexes.
	An effective Hamiltonian demonstrates that a variety of types and numbers of nodal loops 
	will emerge as parameters are varied. 
	This model provides guidance for engineering topological transitions in 
	CaAs$_3$ and related materials by applying external tensile or compressive 
	strains, or by alloying with isovalent atoms on either site.

	We have benefited from comments on the manuscript from A. Essin, from
	discussions of topological aspects with K. Koepernik and J. K\v{u}nes,
	and from T. Siegrist and A. P. Ramirez on CaAs$_3$ samples and transport
	data. The calculations used high 
	performance clusters at the National Supercomputer Center in Guangzhou, and
	resources of the National Energy Research Scientific Computing
	Center (NERSC), a DOE Office of Science User Facility supported by the Office of
	Science of the U.S. Department of Energy under Contract No. DE-AC02-05CH11231.. 
	W.E.P. was supported by the NSF grant DMR-1534719 under the 
	program {\it Designing Materials to Revolutionize and Engineer our Future}. 
	Z. P. Y. and Y. Q. were supported by the National Natural Science Foundation of China (Grant No. 11674030) and the National Key Research and Development Program of China (contract No. 2016YFA0302300) .

\section{CaAs$_3$}
The boundary spectra of nodal loop semimetals is has intriguing aspects.
Without spin-orbit coupling (SOC) the projection of the nodal loop
of accidental degeneracies onto the surface  provides a ``patch''
(an area) in which surface states reside, as opposed to topological
insulators where only isolated surface bands appear. 

\begin{figure}[!hb]
    \includegraphics[width=0.85\columnwidth]{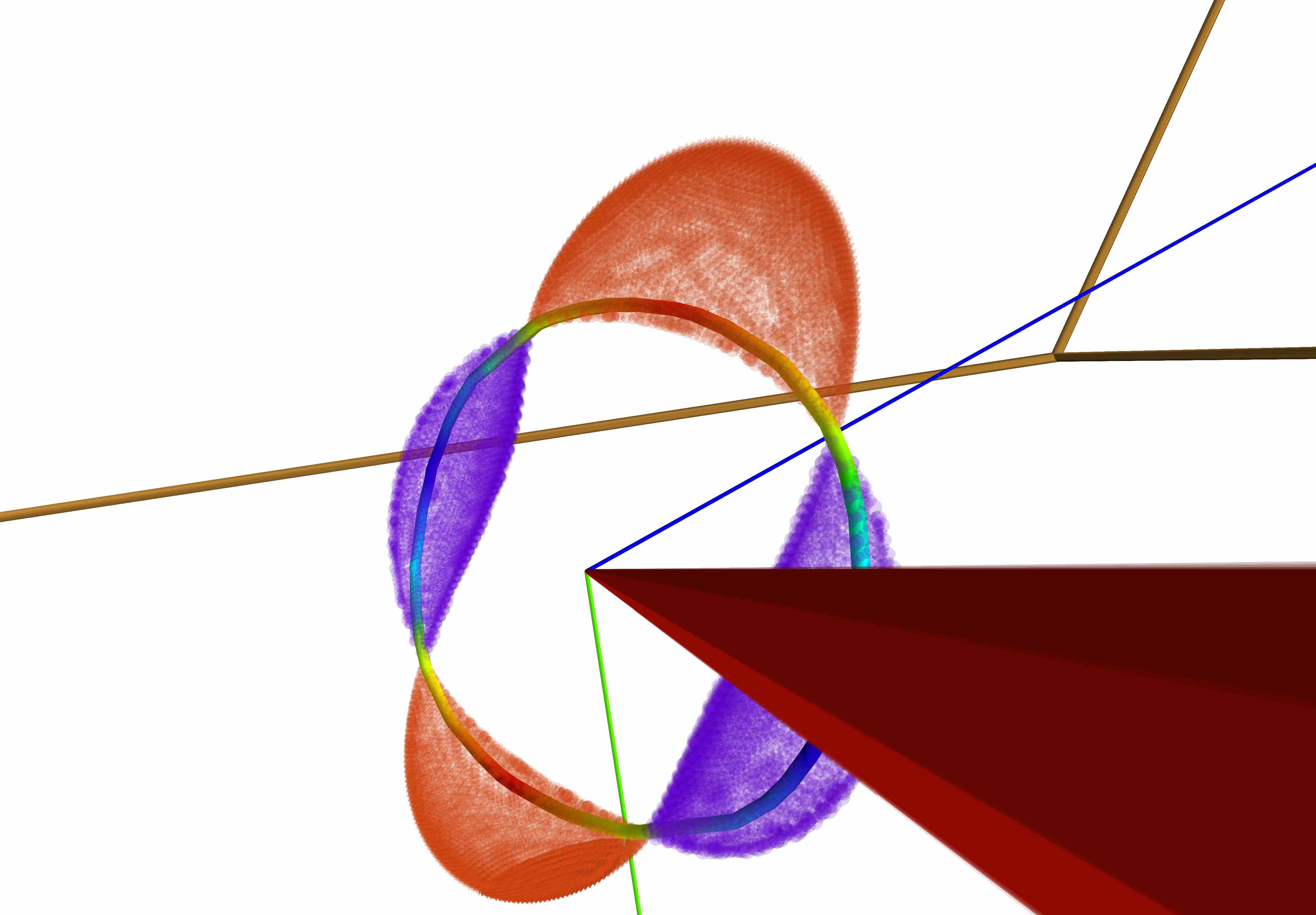}
    \vskip 6mm
    \includegraphics[width=0.85\columnwidth]{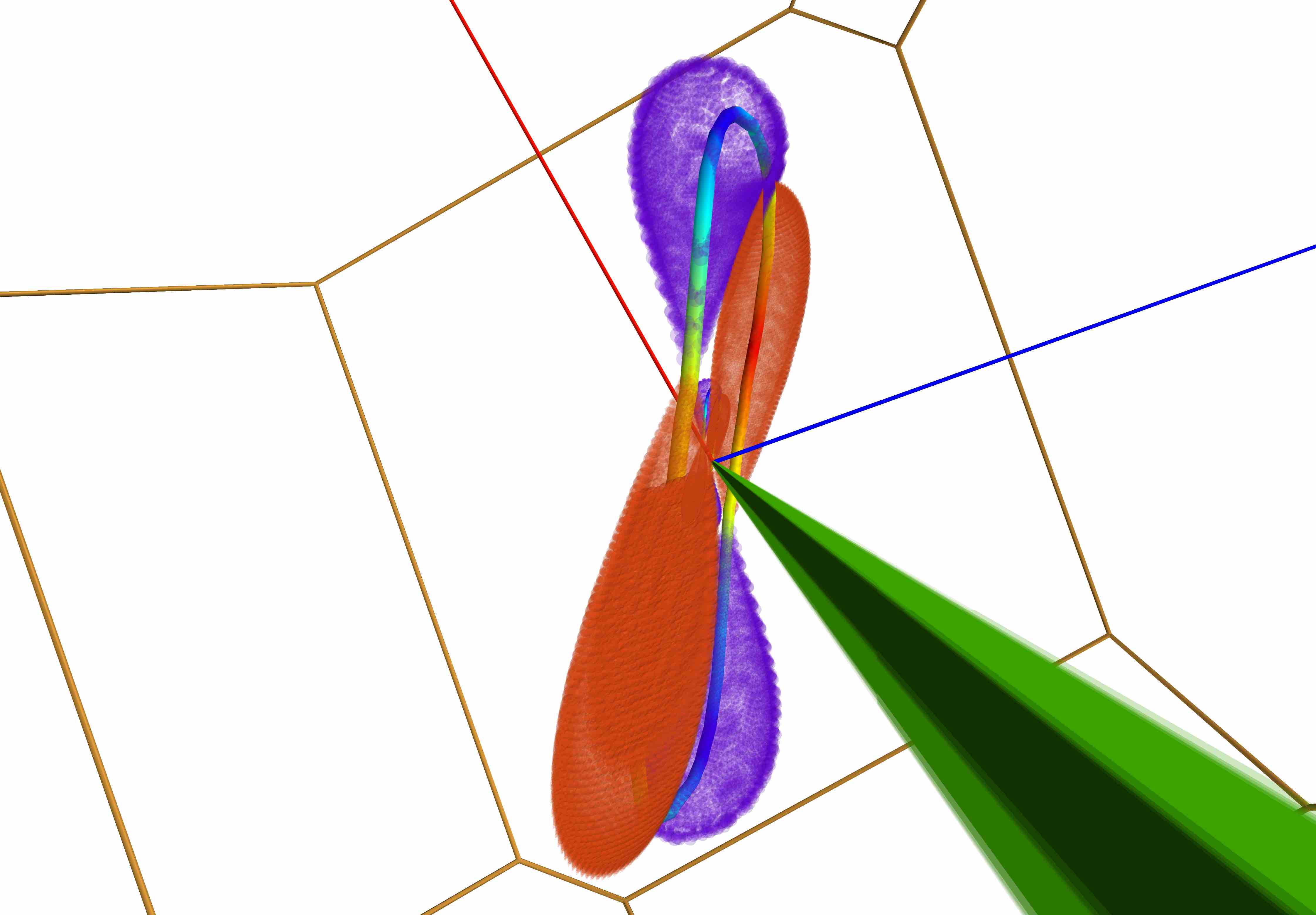}
    \vskip 6mm
    \includegraphics[width=0.85\columnwidth]{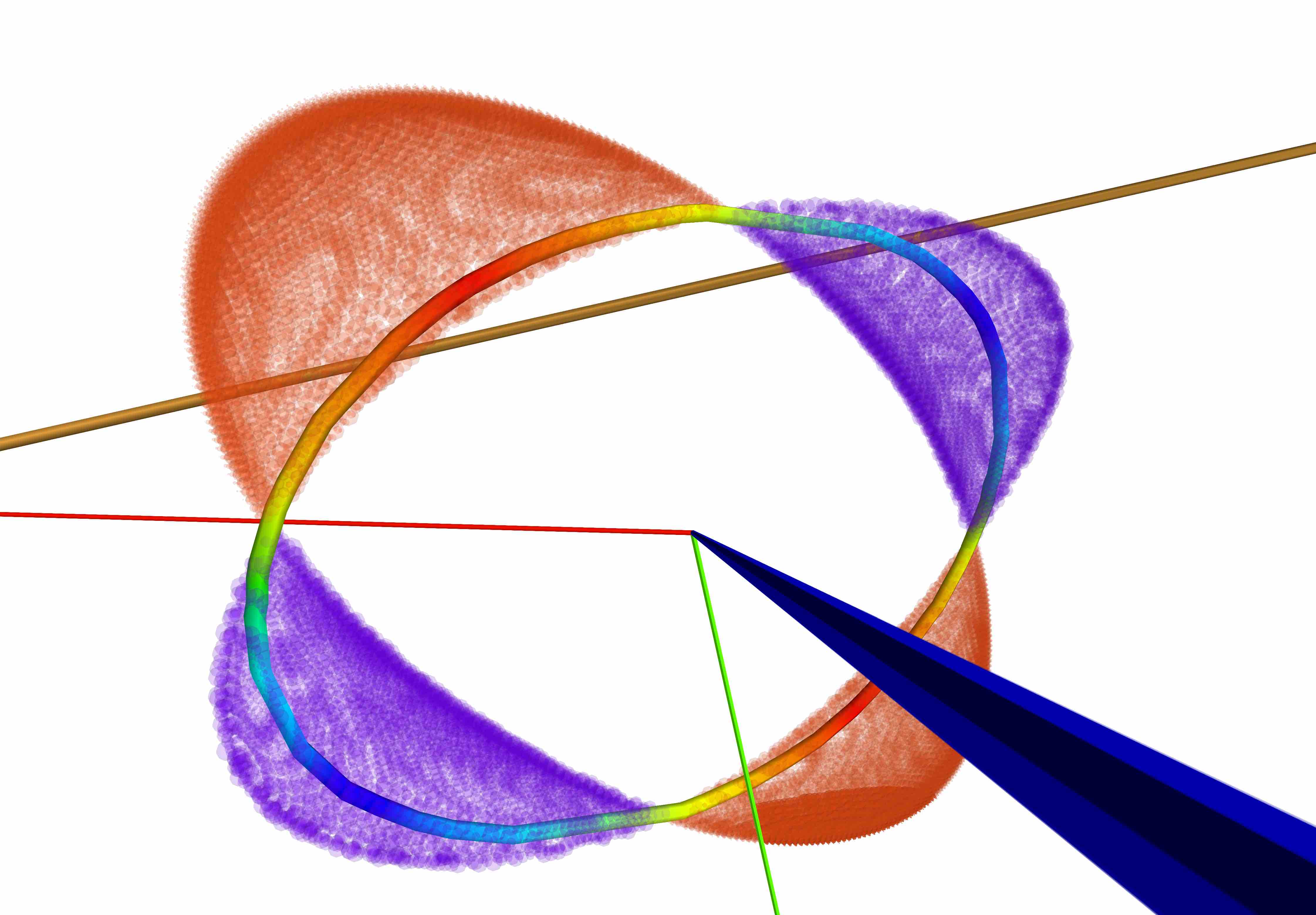}
    \caption{Views of the nodal loop of CaAs$_3$ from approximately
    down the axes, as indicated. This view indicates the shape of
    the patch containing drumhead states for each of the projections
    onto surfaces. The axes extending out of the page (and becoming
    enlarged are: top, $A^*$, middle, $b^*$, bottom, $c^*$.)
     }
    \label{projections}
    \end{figure}

Figure~\ref{projections} provides three views of the nodal loop
(spin0orbit coupling neglected)
and surrounding Fermi surface of CaAs$_3$ that indicates what
the projected patches for topological surface states are like.
Already the loop is small (see main text), and the $b^*$
projection extremely narrow, giving a patch with tiny area.
Figure XX in the main text indicates how spin-orbit coupling
affects the surface spectrum along two special lines, leaving
topological surface state and a topological insulator or
topological semimetal depending on whether a gap is fully
opened, or (indirect band overlap remains.`

\section{The two-band model}
\begin{figure}[!ht]
    \includegraphics[width=0.95\columnwidth]{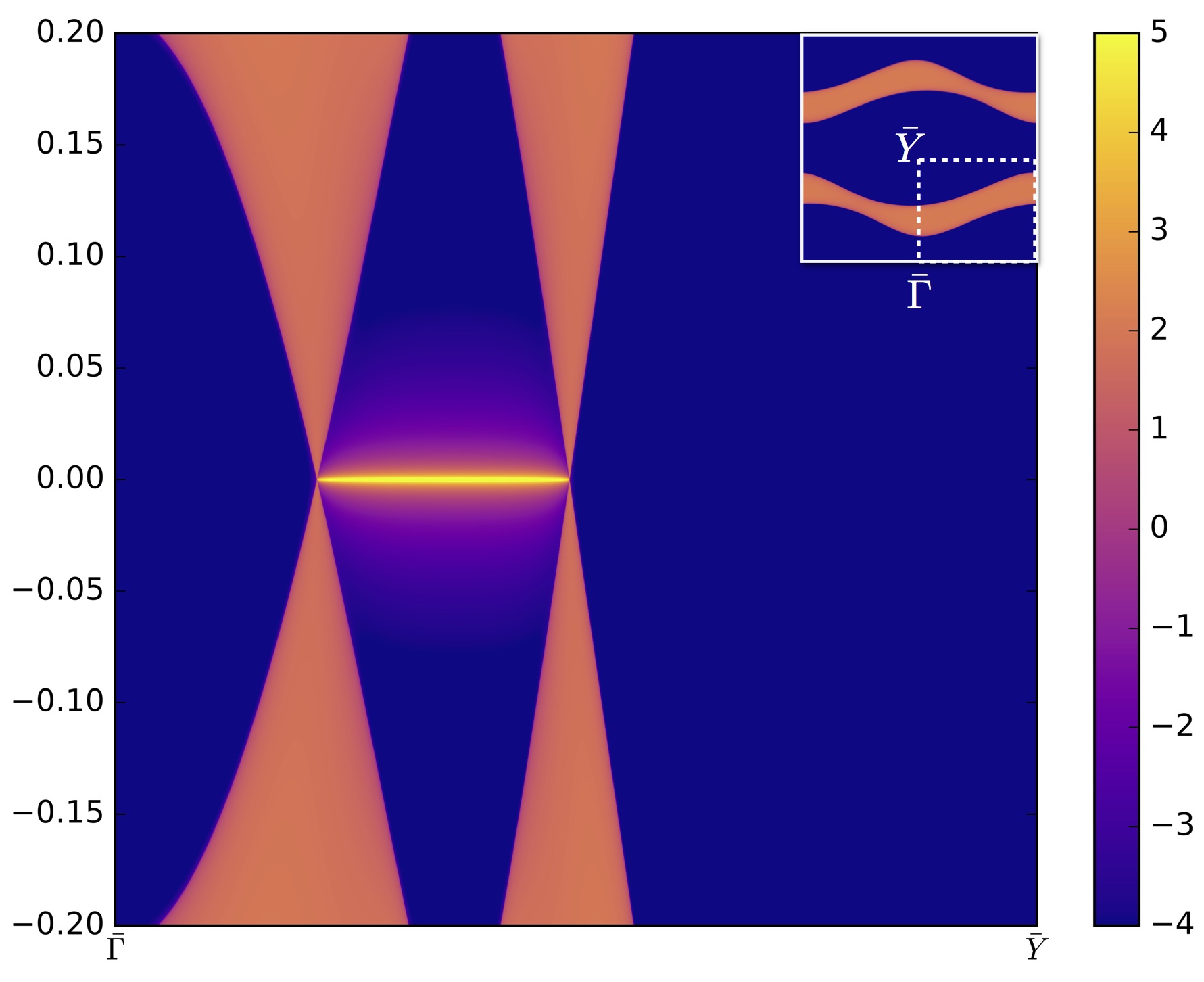}
    \includegraphics[width=0.95\columnwidth]{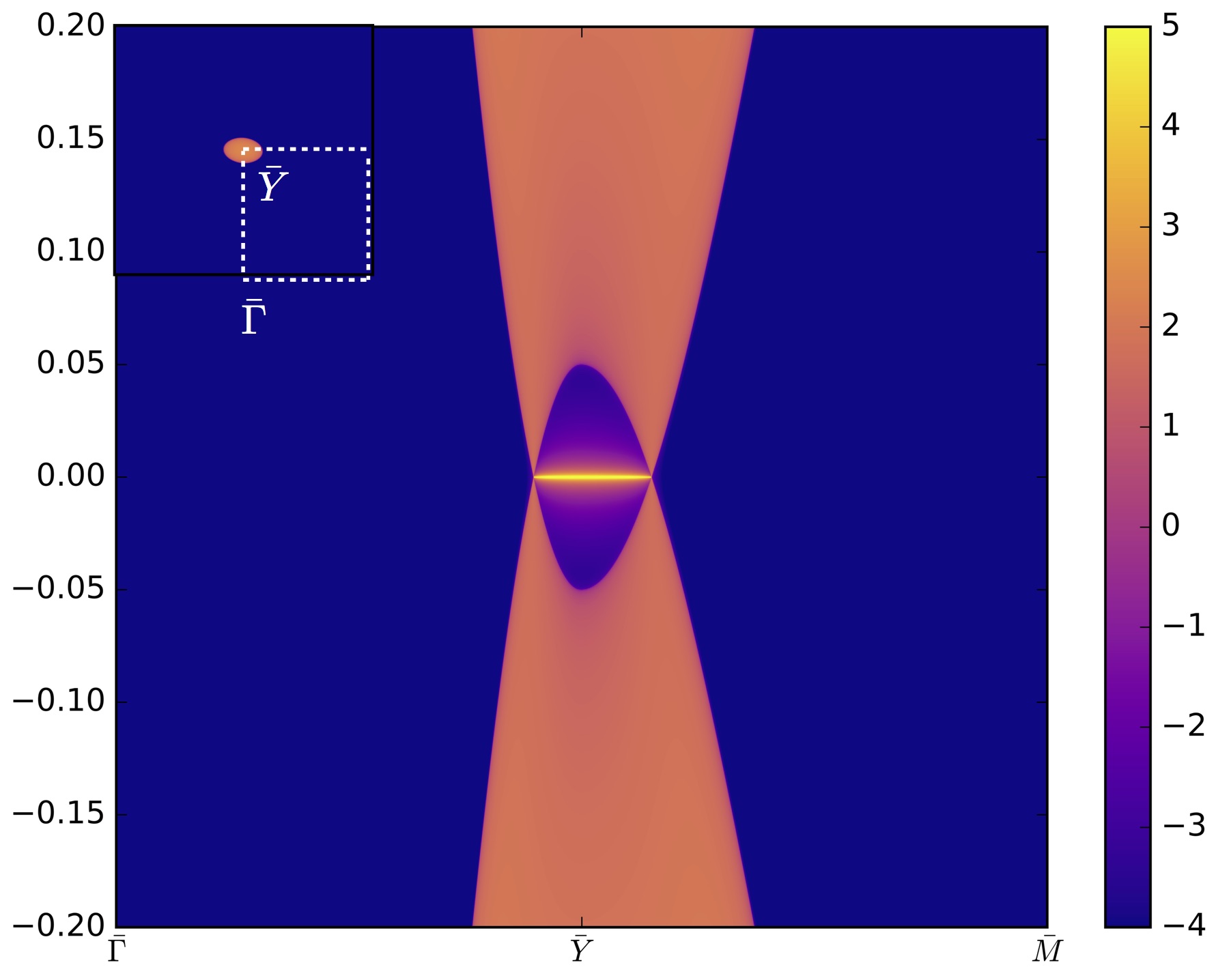}
    \includegraphics[width=1.15\columnwidth]{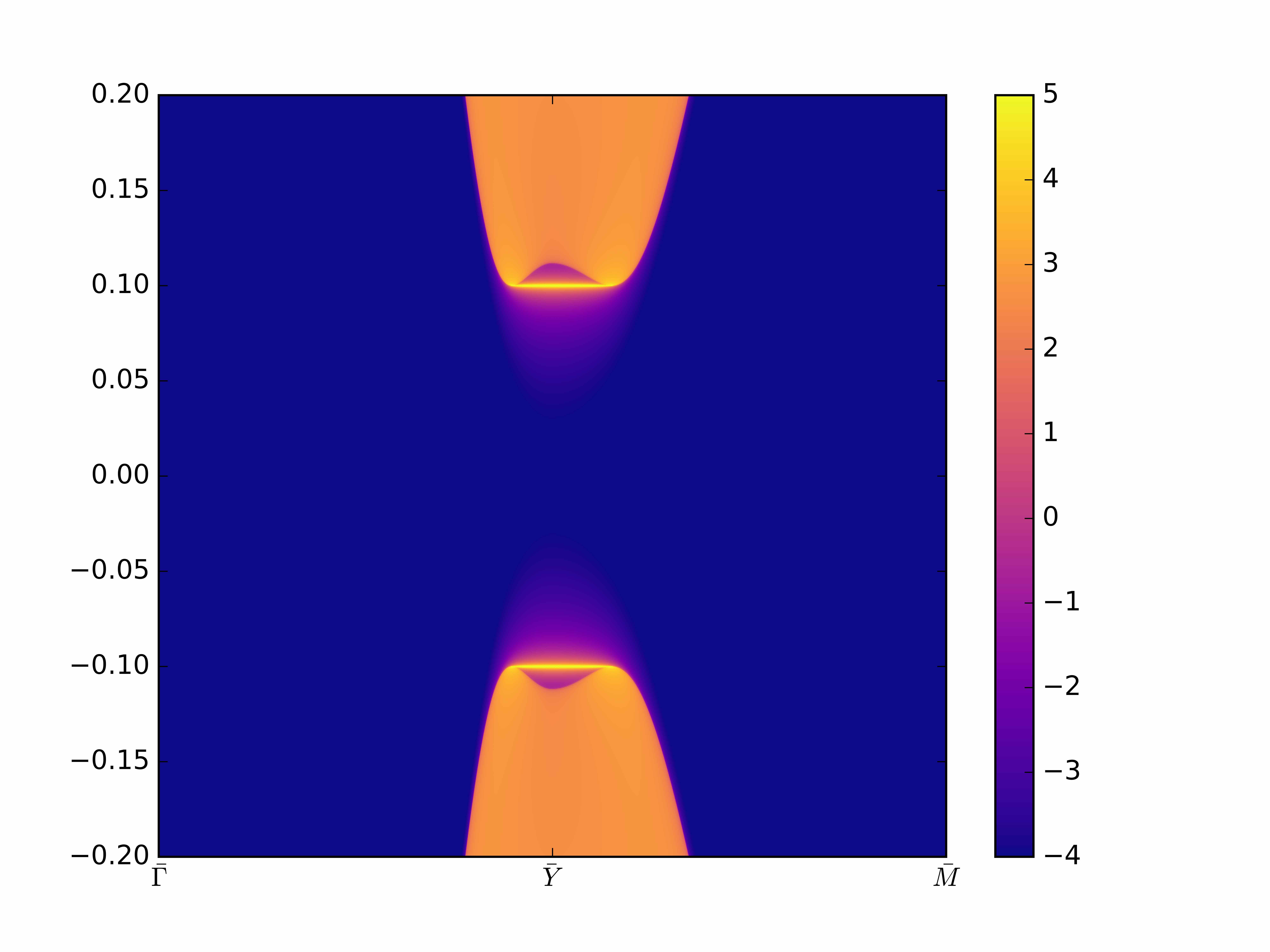}
    \caption{Surface spectrum from the two-band model, showing the surface
  bands and projected bulk DOS along the indicated lines. 
  The top two panels correspond to $m$=1.64, then $m$=0. The inset
  in each case shows the drumhead states lying within the patch 
  defined by the projection of
  the nodal loop onto the surface. 
  See main text for the two-band model and 
  definition of parameters. The bottom panel shows the surface
  spectrum after inclusion of spin-orbit coupling. A simple and
  clear bulk gap is opened and no topological states arise.}
    \label{surfacespectrum}
    \end{figure}

In Fig.~\ref{surfacespectrum} the  spectrum projected on the surface perpendicular to
lattice vector $b$ is shown for the two-band model in the main text,
for two values of the ``mass'' $m$ (the separation of the two bands).
Due to the high (cubic) symmetry of the model, the nodal loop for  
the surface band along the symmetry lines is flat. and the particle-hole
symmetry (evident in the orange projected bulk band structure)
puts the Fermi level at the points of degeneracy along
these lines. 

The insets indicate the surface patches defined by the
projection of the nodal loop onto the surface. For $m$=0with the single
small loop (see Fig. 5 of the main text) the patch is a small ellipse
around $\bar{Y}$.  The $m$=1.64 case is instructive, as it contains
nodal loops that are only closed by the periodicity of the zone.
Unlike the single loop for $m$=0, these nodal lines appear in
inversion symmetry dictated pairs. One pair is crossing roughly
the middle of the zone, whereas the other pair wiggles along the 
top.bottom of the zone. The projections gives two pairs of lines,
with a chirality that results in the patch of drumhead states
choosing to be on one particular side of edge of the patch.
Again, the symmetry of the model leads to little if any variation
in the intensity of the surface drumhead states across the patches.

\end{document}